# High Order Dynamic Mode Decomposition for Mechanical Vibrations and Modal Analysis

**Authors**

Andreas Tuor[1], Nico Canzani[2], Tobias Rüggeberg[3], Stefan Gorenflo[4], Gerd Simons[5]
Bruno Bättig[6], Daniel Iseli[7]

**Main Author**

Andreas Tuor          andre@tuor.net

**Corresponding Author**

Gerd Simons           gerd.simons@fhnw.ch

**Other Authors**

Nico Canzani          nico.canzani@fhnw.ch
Tobias Rüggeberg      tobias.rueggeberg@fhnw.ch
Stefan Gorenflo       stefan.gorenflo@fhnw.ch
Bruno Bättig          bruno.baettig@vibro-consult.ch
Daniel Iseli          daniel.iseli@vibro-consult.ch

**Author details**

[1,2,3,4,5]   Fachhochschule Nordwestschweiz, Institut für Sensorik und Elektronik,
    Klosterzelgstrasse 2, CH-5210 Windisch

[1]     Andre Tuor, not affiliated to FHNW anymore

[6,7]   Vibro-Consult AG, Mess- und Schwingungstechnik
    Stahlrain 6, CH-5200 Brugg

**Abstracts**

In many mechanical, electrical, and general physical systems evolving over time or space, spectral analysis methods as Fast Fourier Transform (FFT), Short Term Fourier Transform (STFT), Power Spectrum Density (PSD) plays a very important role. They allow an extraction of required information content from signals in another base by decomposing it in its spectral components for further processing.

In theory this approach is very powerful, even in some 'simple' or 'not too complicated' practical cases it has proven its utility and efficiency. However, for real-world applications such as mechanical modal analysis of large dimension systems including damping, noise and unpredictable excitation those signals are often so complex that it can be almost impossible to obtain a high-resolution spectral decomposition with these methods due to the time-bandwidth limitation.

In this paper we describe an alternative approach for spectral analysis based on the High Order Dynamical Mode Decomposition (HODMD) and Kernel Density Spectrum (KDS). We will show that this method allows overcoming some limitations of the FFT and may be a promising approach to for a much more precisely the spectral decomposition.

# 1. Introduction

Modal vibration analysis can be performed as modal decomposition of measured displacements, velocities or accelerations. In a classical approach this is generally performed by spectral analysis using FFT/PSD/STFT or similar. They are all based on FFT which has some major drawbacks:

1. The FFT does not take into account the decay of typical vibration signals. In fact, FFT is based on the hypothesis that the analysed signals (more precisely their spectral components) have a constant amplitude over the whole analysed range (the time). The consequence of this assumption is a spread spectrum around the "effective" frequency for decaying oscillations. Adjacent modes are not visible in such spread frequency peaks.
2. The spectral frequency resolution is defined by the number of samples and the sampling frequency ($F_{bin} = F_{sampling}/N_{samples}$) and is "frequency-class truncated". Indeed, if a better resolution is required, the number of samples must be increased accordingly. Moreover, after the spectral decomposition, the energy is defined in a frequency band, losing its numerical instantaneous resolution.
3. The fixed number of samples introduces a windowing effect on the spectrum leading to frequency leakage. Even if this can be attenuated using a suitable window shape, it will spread the spectrum.
4. The obtained spectrum is highly over-dimensioned. For a real signal of N samples, we obtain N/2 frequency bands. One can argue that most of these bands have no or very low energy and can be "filtered" out, but for a damped and noisy signal this is far from being easy. We can still argue that it is possible to apply statistical post-processing on time-evolving spectrum (STFT, Matlab PWELCH), but this is obtained to the detriment of temporal resolution/evolution and can be highly power consuming.
5. Very simple systems may exhibit apparent chaotical behaviour (see the Lorenz attractor). This is even more true for complex systems with several tens, hundreds or even unknown number of modes. It is uncertain that simple FFT based approaches are able to catch the inherent regularities and dimension (number of modes) of such systems.

Several methods such as wavelet, principal order decompositions or phase-space decompositions have been developed to overcome some of those limitations [12], [13], [14], [15], [16], but not all of them together. In this paper, we use a new state-of-the-art alternative and very promising approach called Dynamic Mode Decomposition (DMD) or Higher Order Dynamic Mode Decomposition (HODMD) combined with a Kernel Density Spectrum (KDS). Those algorithms allow us to extract a sparse (dominant) spectral decomposition of measured complex time signals (e.g. accelerations), similar to FFT, but including decay information for each obtained mode without spectrum spread and numerical class rounding. In our case we will apply this method on the vibrations of a railway axle shaft, and it will be shown to be a very powerful method overcoming the previous mentioned restrictions.

## 2. Method (The Dynamic Mode Decomposition and Spectrum generation)

This section will give a brief intuitive summary of the DMD and HODMD decomposition. For more details the reader should consult the references, especially [1], [2], [3], [4], [5] and several videos of lectures [17].

DMD algorithm was first introduced and named by P.J. Schmid in 2008 to identify high-dimensional spatio-temporal coherent structures in (non-linear) fluid flow dynamics. It is a purely data-driven approach, no knowledge of the dynamic of the system or its governing differential equations is required.

To quote [1]:
p. 2: "The method (DMD) can be thought of as an ideal combination of spatial dimensionality-reduction techniques, such as the proper orthogonal decomposition (POD), with Fourier transform in time" and
p. 236: "DMD may be formulated as an algorithm to identify the best-fit linear dynamical system that advances high-dimensional measurements forward in time. In this way, DMD approximates the Koopman operator restricted to the set of direct measurements of the state of a high-dimensional system".

Using an approach developed for non-linear system for "simple" vibration analysis (approximated as linear problem) may sound overturned. But we must bear in mind that we are dealing with complex high-dimensional systems where linearity can be hidden in this complexity or may even not being guaranteed anymore for large displacements.

As pointed out in [4], classical DMD suffers from a very restricting practical aspect: it fails if temporal complexity (N) is larger than spatial complexity (M). Under complexity we understand the number of frequencies or modes. In other words, it is necessary to use HODMD when the number of time-frequencies modes N is large compared to the number of spatial-frequencies modes M. This failing condition rapidly appears if a small number of sensors is used.

To overcome this restriction, references [3], [4] and [5] introduces HODMD as an extension of DMD. It can be understood as a tool to analyse high-dimensional non-linear dynamic systems in their phase-state (or state-space).

In this paper we focus our analysis on HODMD algorithms described in [3] and the related Matlab scripts.

A good resume of DMD algorithm can also be found in [7], and a somewhat more historical intuitive "description" in [8].

### 2.1 The classical Dynamic Mode Decomposition (DMD)

DMD is a purely data-driven algorithm, consisting in a regression of a dataset representing the dynamic of a system which is considered as locally (timely and spatially) linear [2].

The easiest way to describe this approach it is by a simple mathematical description. Let's consider a dynamical system, $\vec{x}$ representing its states (in some cases directly the signals), $t$ the time, $f$ the system itself and $\mu$ the parameter(s). Its dynamics is given by a set of continuous or discrete formulated linear or non-linear first-order coupled ordinary differential equations:

$$\frac{d\vec{x}}{dt} = \vec{f}(\vec{x}, t, \mu) \quad or \quad \overrightarrow{x_{k+1}} = \vec{F}(\overrightarrow{x_k}, \mu) \quad (1)$$

The DMD approach takes an equation-free perspective where the dynamic is unknown, and constructs a locally linear dynamic system formulated by:

$$\frac{d\vec{x}}{dt} = \mathcal{A}\vec{x} \quad or \quad \overrightarrow{x_{k+1}} = A\overrightarrow{x_k} \quad (2)$$

where $\mathcal{A}, A$ are respectively the matrix of the continuous- and discrete-time dynamics.

This formulation relies on the assumption that the system time snapshots lie in the invariant subspace under the action of the Koopman group [9], [10], and therefore all linear operation (especially the time-shift, readers familiar with quantum mechanics will appreciate this fact and should think about the time-shift operator...) applies to it.

The general solution for those systems corresponds to a sum of oscillations (as expected in the vibration analysis we want to perform) multiplied by some components containing the exponential decay. This solution is given by its initial condition $\overrightarrow{x_0}$ and:

$$\vec{x}(t) = \sum_{k=1}^{n} \overrightarrow{\phi_k} \exp(j\omega_k t) b_k = \mathbf{\Phi} \exp(j\mathbf{\Omega}t)\mathbf{b} \quad or \quad \overrightarrow{x_k} = \sum_{j=1}^{r} \overrightarrow{\phi_{kj}} \lambda_j^k b_j = \mathbf{\Phi}\mathbf{\Lambda}^k\mathbf{b} \quad (3)$$

$\overrightarrow{\phi_k}, \mathbf{\Phi}$ and $\omega_k, \mathbf{\Omega}, \lambda_k, \mathbf{\Lambda}$ are respectively the eigenvectors and eigenvalues of the matrix $\mathcal{A}, A$ the coefficients, $b_k$ and $\mathbf{b}$ are the coordinates of the initial conditions $\overrightarrow{x_0}$ in the eigenvector basis and $A = \exp(\mathcal{A}\Delta t)$.

To find the linear approximation of Matrix $A$, m measurement snapshots $X_k$, equispaced in time, are taken. Even if, theoretically, two snapshots $[\overrightarrow{x_k}\ \overrightarrow{x_{k+1}}]$ are enough, the input is overdetermined in order to obtain a better catch of the dynamic of the system. The $X$ measurements and its time shifted version $X'$ are given by:

$$X = \begin{bmatrix} | & | & & | \\ \overrightarrow{x_{t_1}} & \overrightarrow{x_{t_2}} & \cdots & \overrightarrow{x_{t_m}} \\ | & | & & | \end{bmatrix} \quad X' = \begin{bmatrix} | & | & & | \\ \overrightarrow{x_{t_2}} & \overrightarrow{x_{t_3}} & \cdots & \overrightarrow{x_{t_{m+1}}} \\ | & | & & | \end{bmatrix} \quad (4)$$

The dynamical system and the best-fit $\widetilde{A}$ matrix are finally given by († representing the pseudo-inverse):

$$X' \approx AX \quad \Rightarrow \quad \widetilde{A} = X'X^\dagger \quad (5)$$

The DMD algorithm produces a low-rank eigendecomposition of the system described by the matrix $A$. It is obtained by Proper Orthogonal Decomposition (POD) rank-reduction for a reduced order model, and optimally fits the measured signal $\overrightarrow{x_k}$ in a least-square sense so that L$_2$ norm $\|\overrightarrow{x_{k+1}} - A\overrightarrow{x_k}\|_2$ is minimised across all values of $k$.

**2.2 The Higher Order Dynamic Mode Decomposition (HODMD)**

As stated previously and shown in [4], DMD fails if the temporal complexity N is larger than spatial complexity M. The authors of [4] showed that if Eq. (2) leads to Eq. (3) the converse is not true; meaning that a simple linearisation is not enough to represent the whole temporal complexity.

To overcome this limitation, they proved that a more general linear decomposition than (2) given by:

$$\overrightarrow{x_{k+d}} = A_1\overrightarrow{x_k} + A_2\overrightarrow{x_{k+1}} + A_3\overrightarrow{x_{k+2}} + \cdots + A_d\overrightarrow{x_{k+d-1}} \quad k = 1,2,\ldots,K-d \quad (6)$$

can catch the whole temporal complexity of the system. The solutions of Eq. (6) are all in the form of Eq. (3) and all expansions of Eq. (3) satisfy Eq. (6) for appropriate d and $A_i$. Latest is assumed "easily" achievable considering that mechanical systems have a limited number of degrees of freedom, dominant modes and must be L$_2$ bounded. The main requirement is to correctly define the dimension d needed to analyse the system.

Eq. (6) can be reformulated through an enlarged-reduced Koopman matrix $A^*$ to bring it back to a similar formulation as Eq. (2) but in higher dimension:

$$\overrightarrow{x_{k+1}}^* = A^*\overrightarrow{x_k}^* \quad \overrightarrow{x_k}^* = \begin{bmatrix} \overrightarrow{x_k}^* \\ \overrightarrow{x_{k+1}}^* \\ \cdots \\ \overrightarrow{x_{k+d-2}}^* \\ \overrightarrow{x_{k+d-1}}^* \end{bmatrix} \quad A^* = \begin{bmatrix} 0 & I & 0 & \cdots & 0 & 0 \\ 0 & 0 & I & \cdots & 0 & 0 \\ \cdots & \cdots & \cdots & \cdots & \cdots & \cdots \\ 0 & 0 & 0 & \cdots & I & 0 \\ A^*_1 & A^*_2 & A^*_3 & \cdots & A^*_{d-1} & A^*_d \end{bmatrix} \quad (7)$$

Finally, the HODMD algorithm can be resumed in following steps:
1. First dimension reduction (spatial reduction)
   Reduce the high spatial dimension input snapshot matrix by truncated Singular Valued Decomposition (SVD). This step can be omitted if the input spatial dimension is limited to 1 or 2 (real or complex input signal).
2. Second dimension reduction
   Apply the higher order Koopman assumption Eq. (6) by using Eq. (7) to the reduced snapshot matrix and reduce the complexity by truncated SVD decomposition.
3. Compute the reduced HODMD modes, growth rates and frequencies formulated by Eq. (3)
4. Compute the HODMD modes amplitudes
   Optionally, mode reduction can be performed by truncated amplitude SVD decomposition
5. Compute the HODMD expansion for the original snapshots for an error estimation

In this paper we have used and adapted the MATLAB scripts provided by [3].

As pointed out, the first step of the algorithm can be omitted if the spatial dimension of the input signal is limited to one or two input signals (sensors).

The truncation limit for SVD reduction can be defined by several ways, especially:
1. Use a standard SVD error estimation defined by a tolerance, removing all $\sigma_i < \varepsilon\sigma_1$, $0 < \varepsilon < 1$,
2. Define the number of dominant modes N ($\sigma_1 \cdots \sigma_N$)
3. Calculate an optimal hard threshold using the approach described in [6]

Generally, we use the third approach for real application noisy measurements and the first for simulations.

As this subject is too complex to be described in more details here, we strongly suggest reading [1], [2], [3], [4] and [5].

## 2.3 FFT-HODMD comparisons of simulated signals

To compare FFT and HODMD, signals and spectrums respectively frequencies are first compared on simulations. The input signals are modelled by a sum of exponentially decaying oscillations having different frequencies and damping factors (all amplitudes are set to 1 for simplicity).

$$S_i(t) = \sum_{k=1}^{i} exp(-tD_k) sin(\omega_k t) = \sum_{k=1}^{i} exp\left(\frac{t}{\tau_k}\right) sin(\omega_k t) \quad (8)$$

where $\omega_k = 2\pi f_k$ represent the angular frequencies and $D_k$ the damping factor (later we also use $\tau_k = -1/D_k$)
The first case compares the extractions of a single decaying oscillation ($i = 1, f_1 = 2000\ Hz, D_1 = 80s^{-1}$), the second signal is composed of tree decaying oscillations ($i = 3, f_1 = 2008\ Hz, D_1 = 50s^{-1}, f_2 = 1992\ Hz, D_2 = 80s^{-1}, f_3 = 1800\ Hz, D_3 = 100s^{-1}$), and the third signal is composed of 8 decaying oscillations with random frequencies and decaying factors. The sampling time is 40uS (F$_S$=25kHz) and the FFT sample length is 64k (2^16 samples, giving a frequency bin resolution of 0.38 Hz).

The plot of the signals and its HODMD based reconstruction as well as the FFT and the HODMD extracted frequencies are represented on **Fig.1**. In all cases HODMD reconstructed signals (based on extracted frequencies, amplitudes and damping values) perfectly fits the original signal (relative RMS and MAX errors are strictly 0), and the extracted frequency, damping and amplitudes exactly match the defined parameters.

It may also be noted that if for the single decaying oscillation, the peak of the FFT is exactly located at the defined and extracted HODMD frequency. The second FFT shows a first peak at about 1797 Hz instead of 1800 Hz and a second peak at about 2009 Hz instead of 2008 Hz. Moreover, the intermediate frequency can hardly be extracted due to the superposition of spread spectrums. On the third signal only 2 peaks can be observed on the FFT spectrum, neither been extracted exactly. The HODMD frequencies, on the other hand, are extracted with an error lower than 0.1Hz.

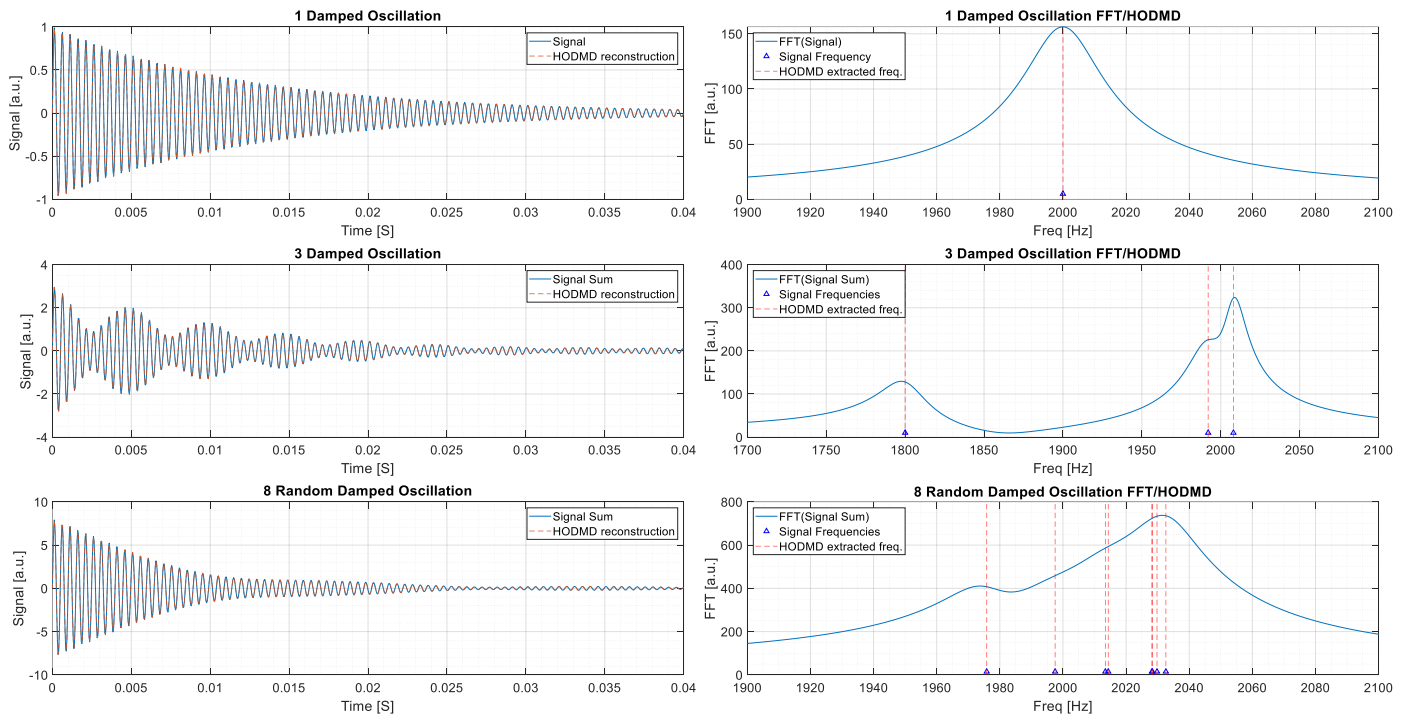

**Fig.1** Left side: signal and reconstructed signal, right: FFT spectrum, HODMD extracted frequencies (dashed vertical lines) and signal used frequencies (triangle). Top: Single decaying oscillation, middle: superposition of 3 decaying oscillations, bottom: superposition of 8 (random frequencies and decaying factors) decaying oscillations.

## 2.4 Spectrum generation from HODMD extractions

Based on the previous HODMD extracted modes, we are able to reconstruct different types of Kernel Density Spectrums (KDS) based on adequate Kernel Density Estimators (KDE). We may note that for simple and nice-looking signals this has only little utility. But we will see in the next chapter that for real signals submitted to unpredictable excitation and noise it will allow us to create sharp high-resolution spectrums.

On the contrary to FFT delivering a class-truncated (easy to represent and analyse) spectral decomposition, HODMD delivers a sparse discrete frequency list of modes and their associated amplitudes and decay factors. Those frequencies live in $\mathbb{R}$, are not rounded nor class-sorted and can be seen as a list of Dirac pulses in the frequency domain (see dashed vertical lines on **Fig.1** right).

To compare HODMD "spectrum" with FFT or PSD we use a Kernel Density Estimator (KDE) to estimate the probability density and to create a Kernel Density Spectrum (KDS) [11]. This spectrum is build using Eq. (9) where $n$ represents the number of samples (number of extracted frequencies), $F$ is the independent parameter (frequency) and $K$ the kernel function. $A_i$, $\tau_i$ and $F_i$ represent the amplitude, time (1/decay) constant and frequency of each mode extracted, and $h$ is a kernel parameter (generally smoothing parameter, also sometimes called bandwidth).

$$KDS(F) = \frac{1}{n}\sum_{k=1}^{n} K\big((F - F_k), A_k, \tau_k, h\big) \quad (9)$$

Different kernels can be used, hereafter we will focus on a Gaussian for the frequency density and a Lorenz to represent the damped oscillators.

a. Gaussian kernel

For the gaussian kernel, we replace each frequency Dirac pulse by a gaussian around the mode frequency:

$$KDS_G(F) = \frac{1}{n}\sum_{k=1}^{n} f(A_k, \tau_k) \cdot exp\left(\frac{-1}{2}\left(\frac{F - F_k}{h}\right)^2\right) \quad (10)$$

The pure density spectrum is calculated using $f(A_k, \tau_k) = 1$ as shown in **Fig.2**. A weighted spectrum can be obtained using a weighting function of $A_i$ and $\tau_i$. The parameter h is chosen to define a bandwidth (corresponding to the gaussian standard deviation), or the frequency-related time-constant. By using $f = A_k^2$, we will obtain a spectrum displaying the power per frequency of each mode which can be compared to PSD, using $f = \tau_k$ will show which frequencies has the lowest decay rate.

The main advantage of the gaussian KDS is its versatility concerning its resolution. The spectrum can be easily tuned by setting $h$ and the range resolution (which should be smaller than $h$) to represent a specific problem. It may be noted for example, that on the lowest figure the two modes around 2014Hz and 2028Hz can be clearly separated (h=0.05).

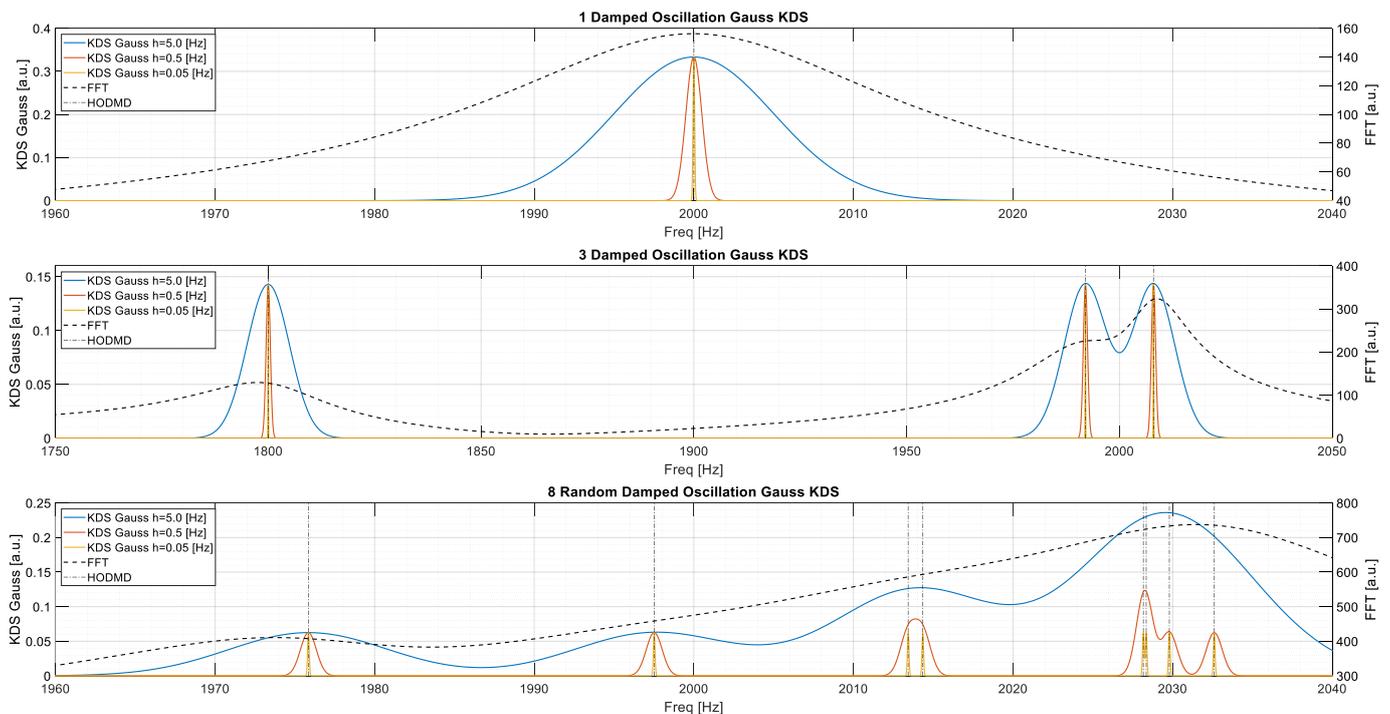

**Fig.2** FFT and Gaussian kernel density spectrum with $f = 1$ and h=0.05, 0.5 and 5.0 Hz applied on HODMD extracted modes. Top to bottom: signal composed by 1, 3 and 8 damped oscillations. Note the mode separation and combination of very near frequencies on the bottom figure.

### b. Lorenz kernel

The Lorenz kernel has its origin in the Fourier transform of a damping oscillation:

$$f(t) = A_k e^{-\left(\frac{1}{\tau_k} - i\omega_k\right)t} \Rightarrow F(\omega) = \frac{A_k}{2\pi} \int_0^\infty e^{-\left(\frac{1}{\tau_k} + i(\omega - \omega_k)\right)t} dt = \frac{A_k}{2\pi} \frac{e^{i \cdot atan[\tau_k(\omega - \omega_k)]}}{\sqrt{\frac{1}{\tau_k^2} + (\omega - \omega_k)^2}} \quad (11)$$

Here we define the Lorenz KDS as:

$$KDS_L(F) = \frac{1}{n} \sum_{k=1}^{n} \frac{\sqrt{h} A_k \tau_k}{\sqrt{1 + 4\pi^2 h \tau_k^2 (F - F_k)^2}} \quad (12)$$

We also included a similar parameter h; it sharps the spectrum when it increases (on the contrary to the gauss kernel). By increasing this parameter, we increase artificially the time during which the oscillation is present. This also brings out the different modes present in the signal. In the **Fig.3** the term $\sqrt{h}$ of equation (12) on the numerator is set to 1 to obtain a better plot scaling on the y axis.

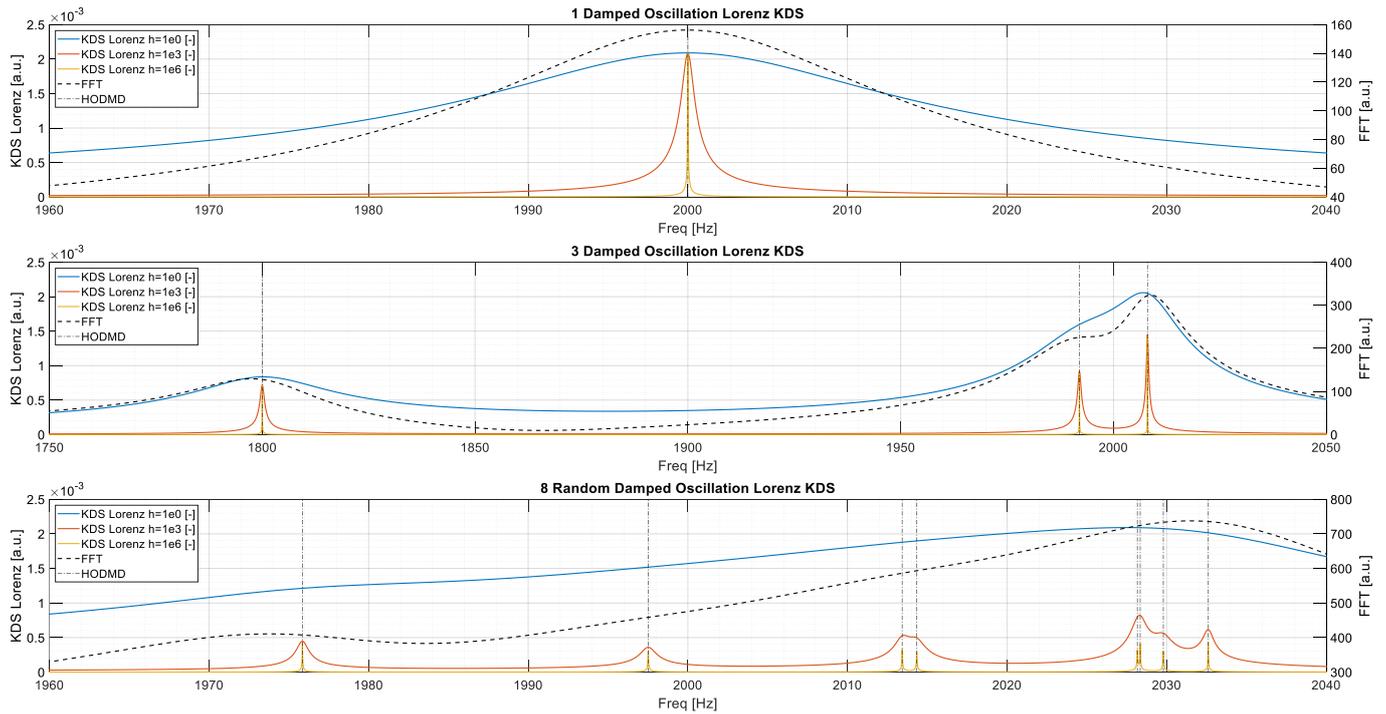

**Fig.3** FFT and Lorenz kernel density spectrum with h=1e0, 1e3 and 1e6 applied on HODMD extracted modes. Top to bottom: signal composed by 1, 3 and 8 damped oscillations. Note the mode separation and combination of very near frequencies on the bottom figure.

## 3. Results and discussions (HODMD applied to mechanical vibrations and modal analysis)

At this point, we will apply the HODMD to perform a spectral analysis on signals delivered by acceleration sensor(s) placed on a railway axle represented in **Fig.4**. Measurements can be obtained in stillstand conditions from sensors placed on the axle or a sensor holder on the axel end with a hammer excitation, or from sensors on the axle end only during drive condition and an excitation from the rail. Hereafter we will look on both approaches.

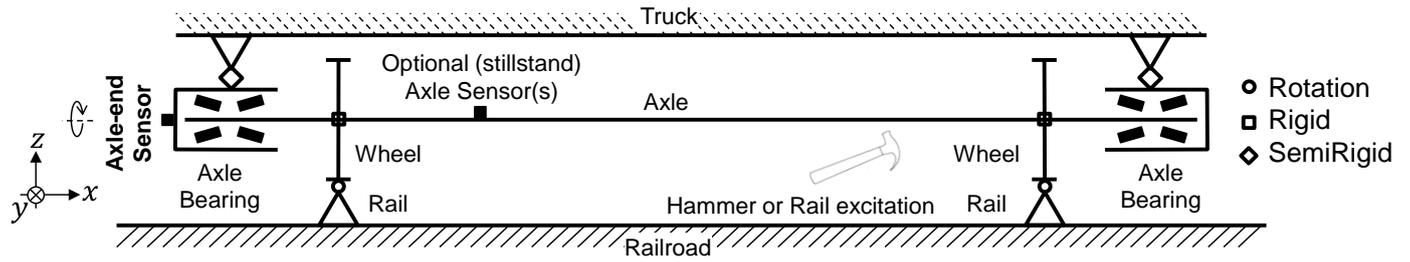

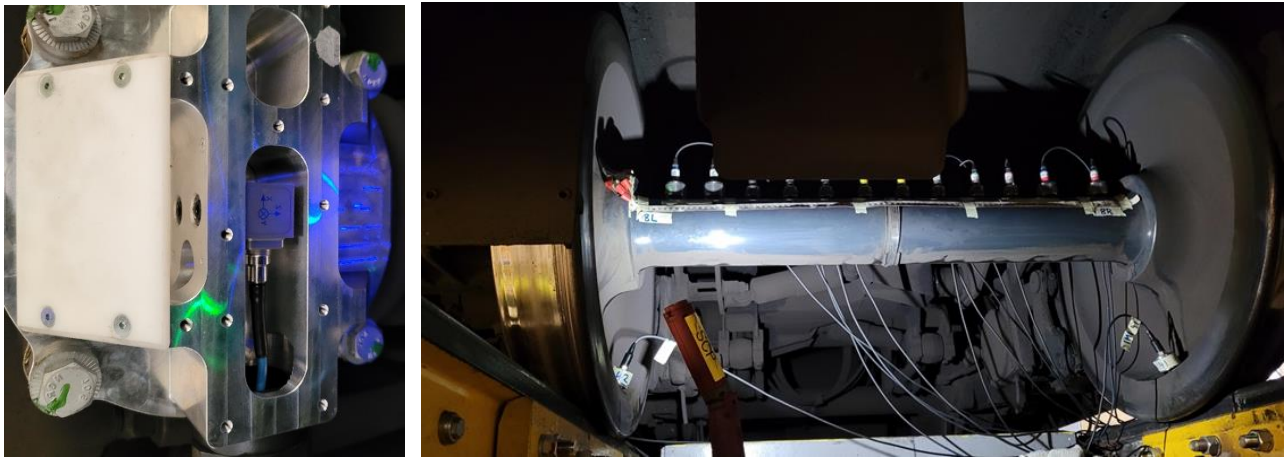

**Fig.4** Schematic drawing of a railway axel and mounted accelerometer sensor(s) and example of mounted sensors on axel end (bottom left) and on the axel (bottom right).

### 3.1 HODMD applied on axle sensor in stillstand excited with hammer hits

Sensors placed directly on the axel end and submitted to hammer hits excitations are less submitted to environmental disturbance and noise. In **Fig.5** we show the Z acceleration modes for 245 hammer hits and a zoom in the first hit.

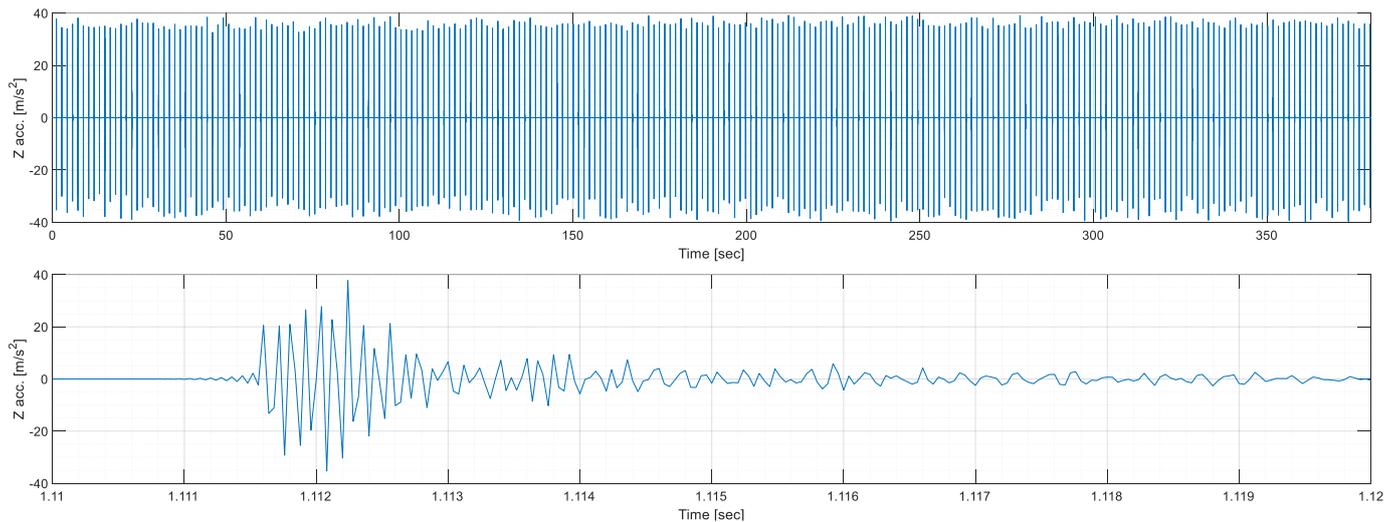

**Fig.5** Z acceleration on axle sensor measured with hammer hit excitation. Top: all hits, bottom: zoom on first hit

A.  HODMD decomposition on a single hammer hit

For each hammer hit 2^10 samples are taken and decomposed with the HODMD procedure as shown in **Fig.6**.

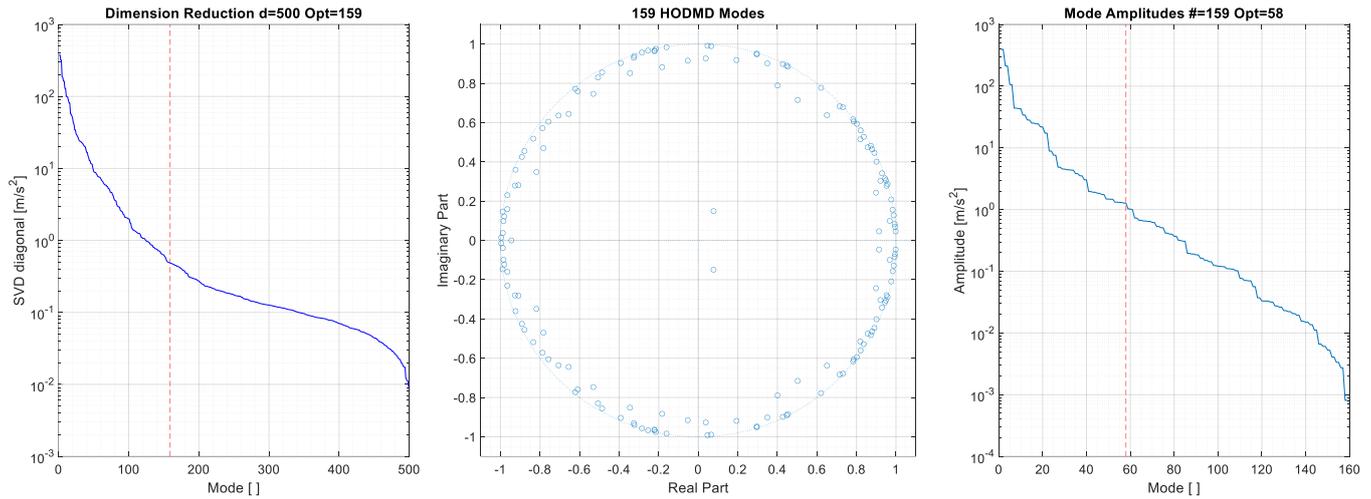

**Fig.6** HODMD modes extraction (see [6] for details). From left to right: second dimension reduction of HODMD algorithm (step 2), modes calculations HODMD algorithm (step 3), amplitude calculation HODMD algorithm (step 4). Vertical dashed lines represent the optimal truncation described in [6].

Extracted results are represented in **Fig.7**, we can observe that the reconstructed signal and its PSD nearly perfectly fits the original.

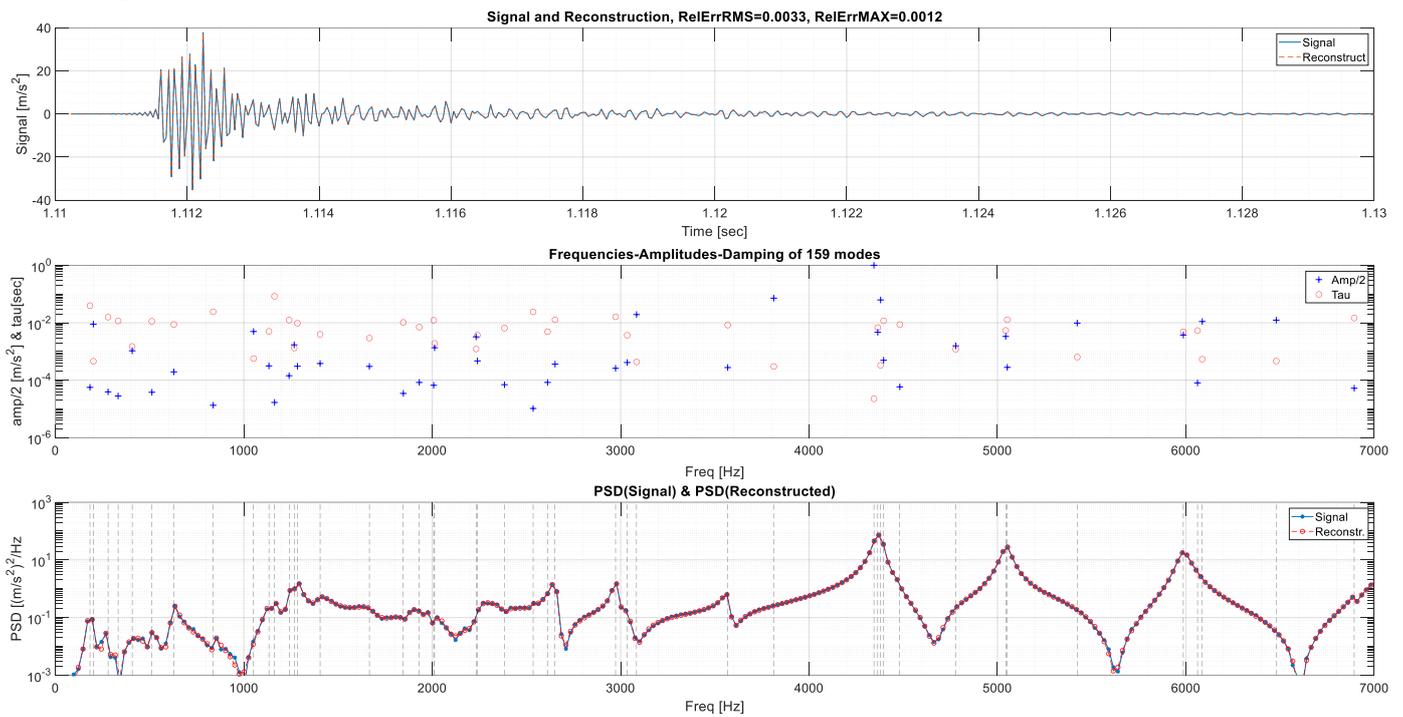

**Fig.7** Axle sensor signal of a single hammer hit and HODMD decomposition. Top: original and reconstructed signal, middle: extracted amplitudes and time constants versus mode frequency, bottom: PSDs of the signal and the reconstructed signal, the dashed vertical lines representing the extracted HODMD frequencies.

B. HODMD decomposition of each hammer hit

For each hammer this decomposition procedure is repeated and the extracted frequencies are represented on **Fig.8**. We can distinguish groups of horizontal dots having similar frequencies, e.g. just above 3500 Hz in the bottom figure, but also spurious frequencies subjected to unpredictable pattern. In this situation, the KDS approach to reconstruct a representative spectrum will show its utility.

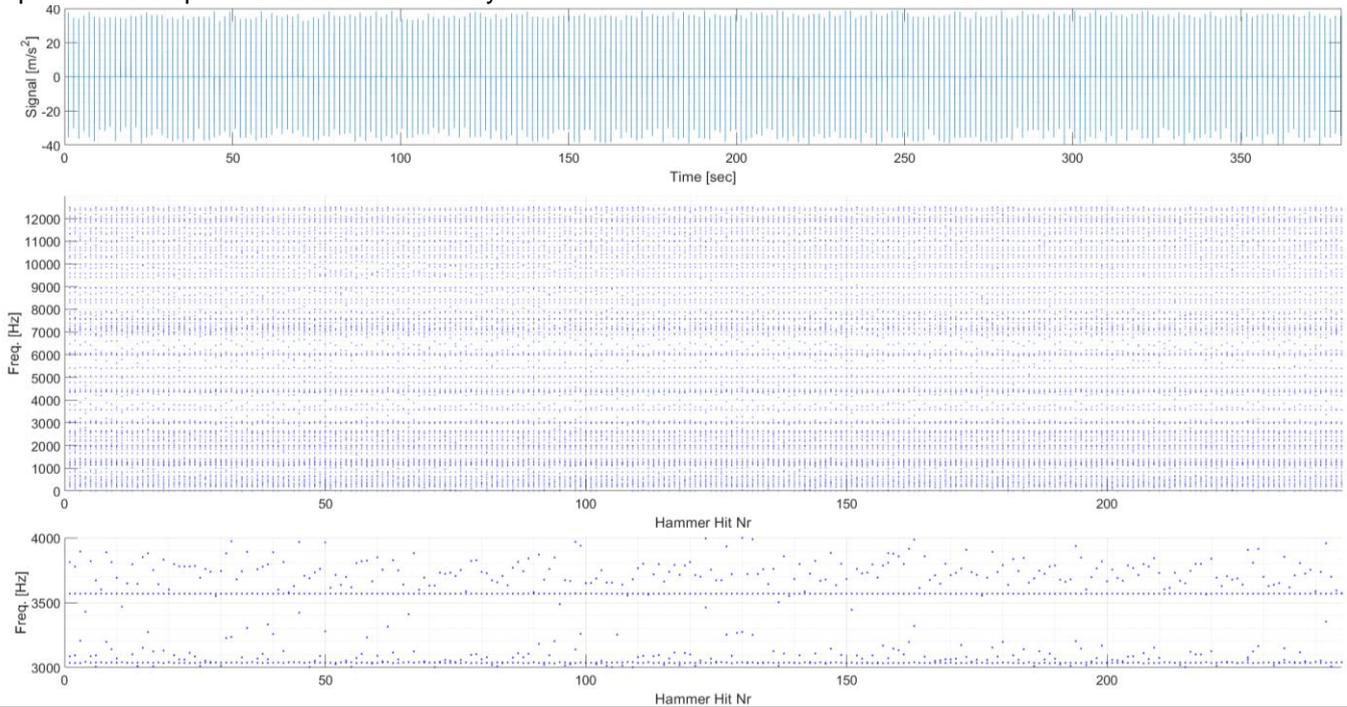

**Fig.8** HODMD decomposition of each hammer hit. Sample length and frequency of each decomposition is 2^10 and 25 kHz. Top: input signal, middle: extraction of frequencies for each hit, bottom: zoom in a frequency range.

C. Spektrum reconstruction using KDS

Based on previous HODMD extraction, we create Gaussian and Lorentzian KDS represented respectively on **Fig.9** and **Fig.10** with different parameters h, both representing a complementary point of view.

For the Gaussian KDS (**Fig.9**), a large bandwidth parameter h filters the different extracted HODMD frequencies and point out which appears frequently, and by decreasing h the different components appear more distinctively until a point where nearly each one is uniquely defined and the spectrum looks very noisy.

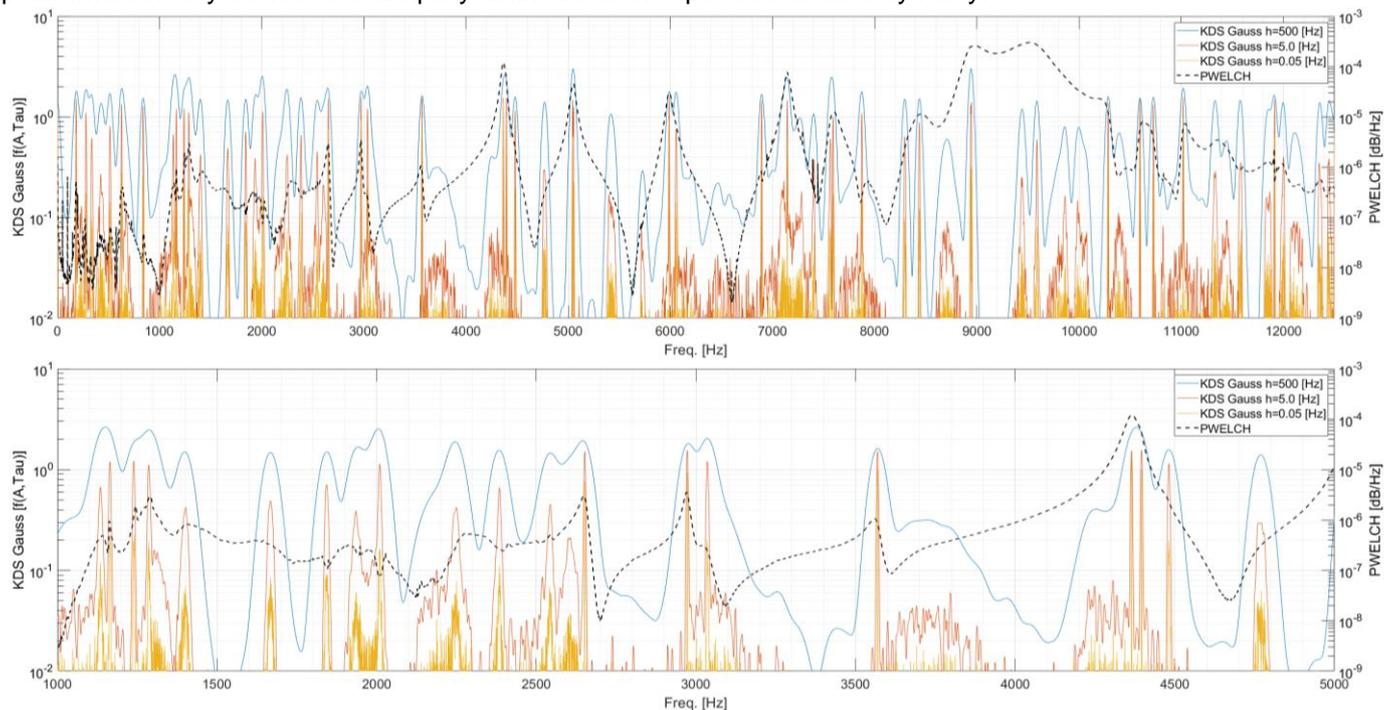

**Fig.9** Comparison of Gaussian KDS with different parameter values of all hammer hits and MATLAB's PWELCH. Top: full spectrum, bottom: frequency zoom

The Lorentzian KDS (**Fig.10**) takes the opposite direction for h. For a small parameter some ''energy humps'' can be observed at frequencies where the PWELCH shows its energy peaks, and by increasing h the different dominant components appear progressively. Note that in some frequency ranges a lot of components rises up in a dense form, and for some frequencies verry sharp spikes popup. For very high h the spectrum has a noisy aspect again.

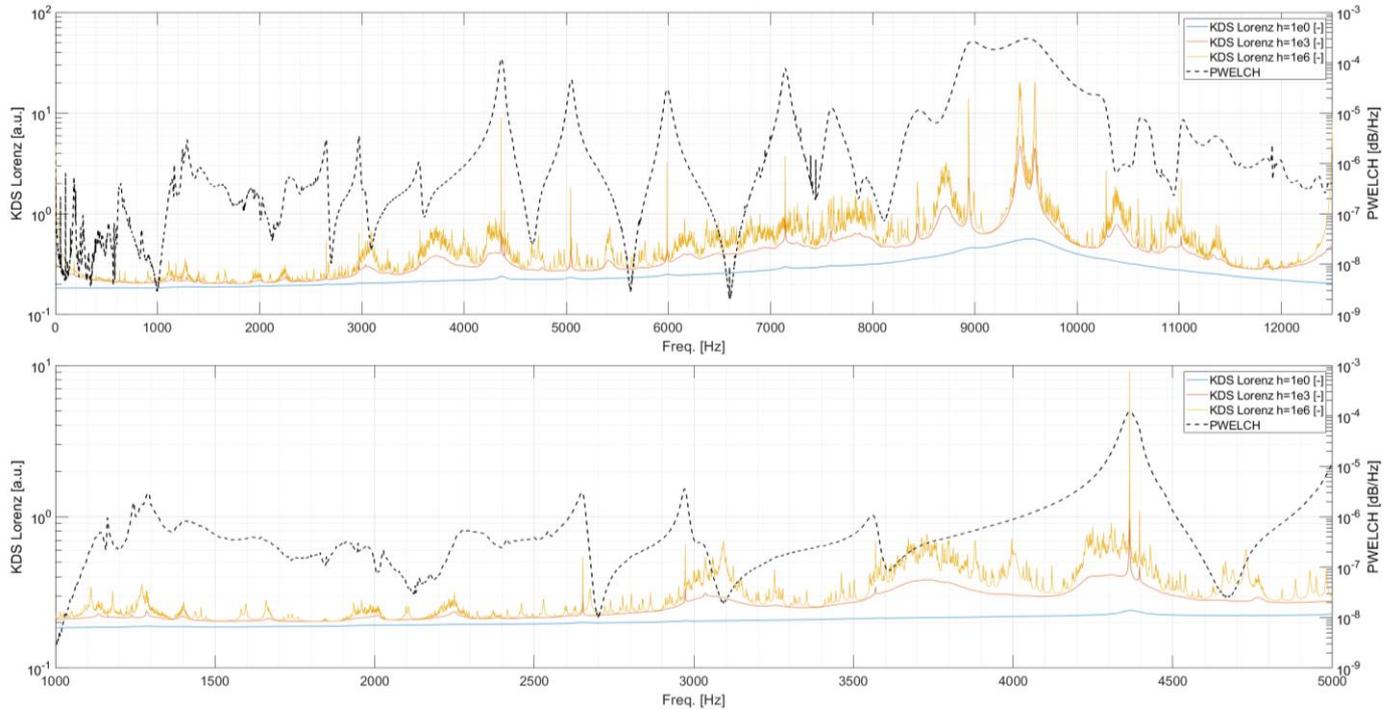

**Fig.10** Comparison of Lorentzian KDS with different parameter values of all hammer hits and MATLAB's PWELCH. Top: full spectrum, bottom: frequency zoom

**3.2 HODMD applied on sensor-holder measurements during operation**

Real-life signals are much noisier than simulations and stillstand measurements. In this chapter we will merely compare HODMD with PSD for a complex noisy signal obtained during operation.
A typical acceleration measurement during operation is shown in **Fig.11**, where the top plot shows the time signal for a period of 600 seconds, the middle plot shows a zoom of 2 seconds, we see two different excitations probably obtained by the railroad rail-transitions, whereas on the bottom plot we see the portion of 0.4 seconds of the signal that we will use for the extraction.

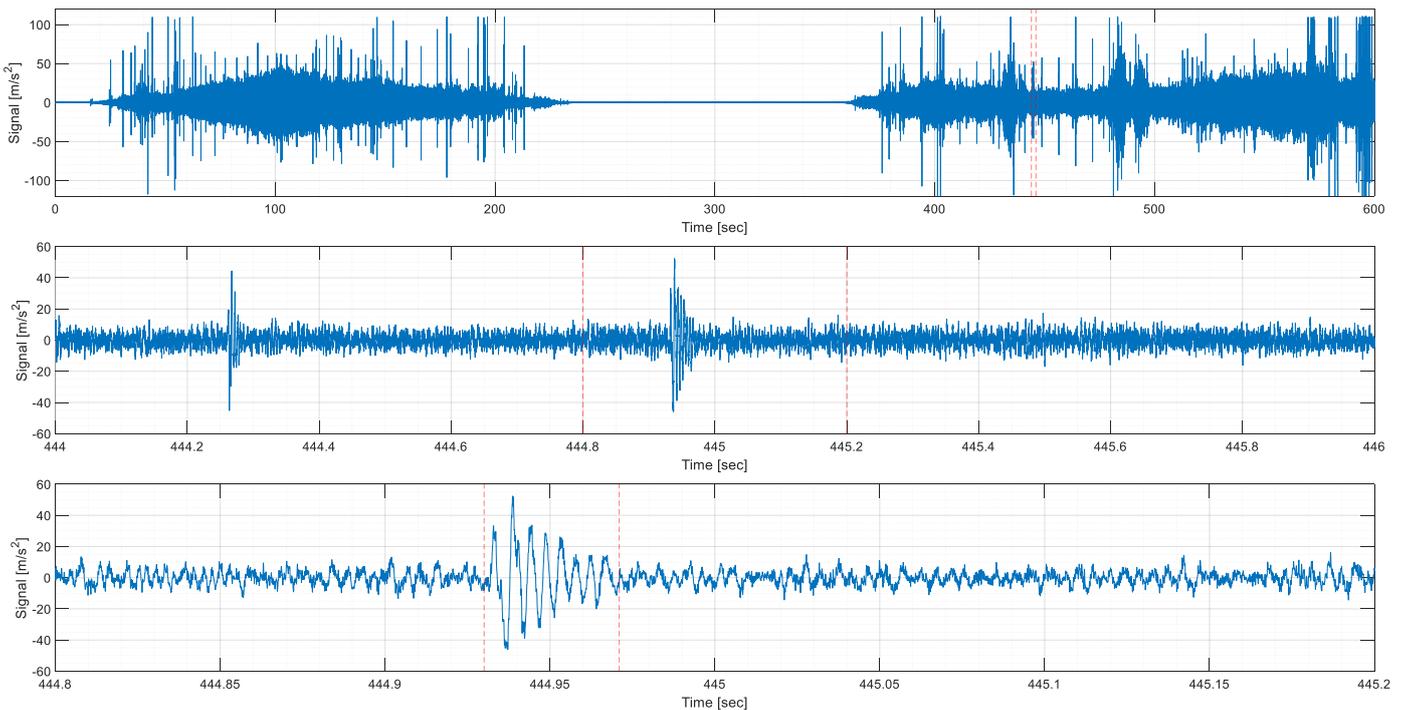

**Fig.11** Signal delivered by an acceleration sensor mounted on a railway axle end (see **Fig.4**). during a ride. Top to bottom: 600 second of delivered sensor signal, zoom of 2 seconds, zoom of 0.4 seconds used below.

First, HODMD algorithm is applied to the region defined by the red marked region on **Fig.11** bottom. As mentioned in chapter 2.2, the dimension of the HODMD extraction space must be chosen properly before.

Previous FEM analysis of the mechanical system pointed out that we may expect up to 200 modes. As those appears in pairs (cf. complex Z plane on **Fig.12** where positive and negative frequencies appear, both frequencies are complex conjugate for a real signal), up to 400 complementary ''modes'' may be needed to represent this system, leading us to use an extraction space d=500. To be able to shift the input measurement dataset d times, the number of snapshots needed is therefore defined by $K > 2d$, for this example we used K=$2^{10}$ timestamps.

The mode decomposition of a single spatial dimension and time snapshots of $2^{10}$ measurement results is shown in **Fig. 12**. The decomposition is reduced to 152 conjugated modes, amplitude decomposition reduction is not used.

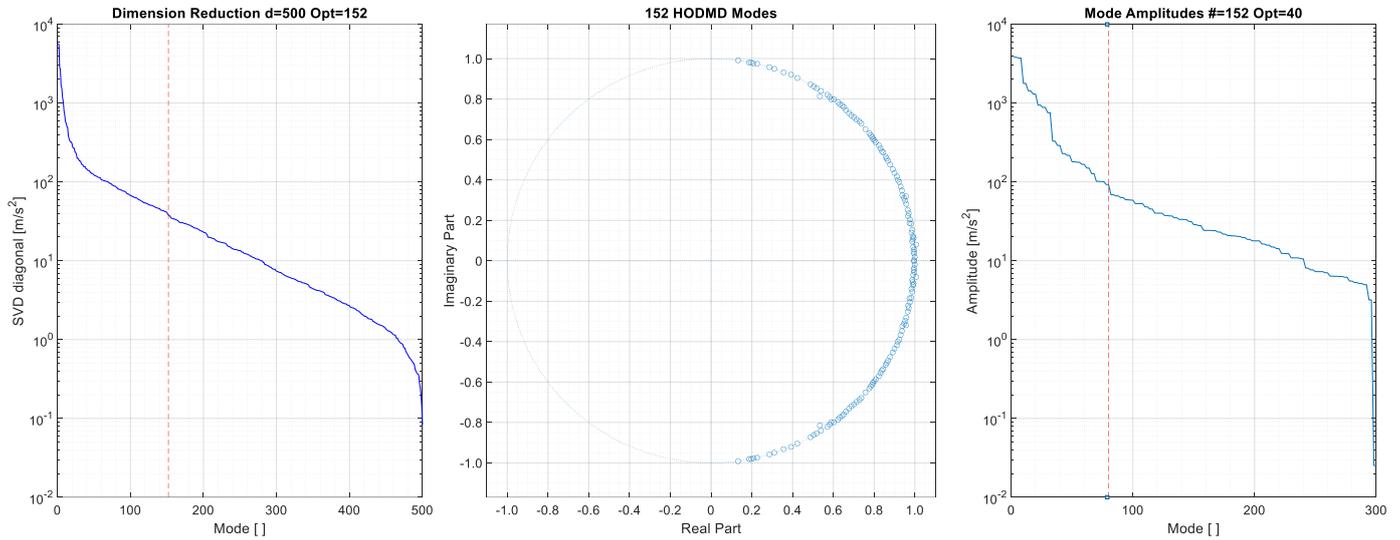

**Fig.12** HODMD modes extraction (see [6]). From left to right: second dimension reduction of HODMD algorithm step 2, modes calculations HODMD algorithm step 3, amplitude calculation HODMD algorithm step 4.

Based on this decomposition, the reconstructed signal, the different modes amplitudes and time constants and PSDs of the original and the reconstructed signal are represented on **Fig.13**. The HODMD reconstruction error remains very small and the spectrum, giving its resolution of approx. 24.4 Hz due to the small number of samples, looks very good.

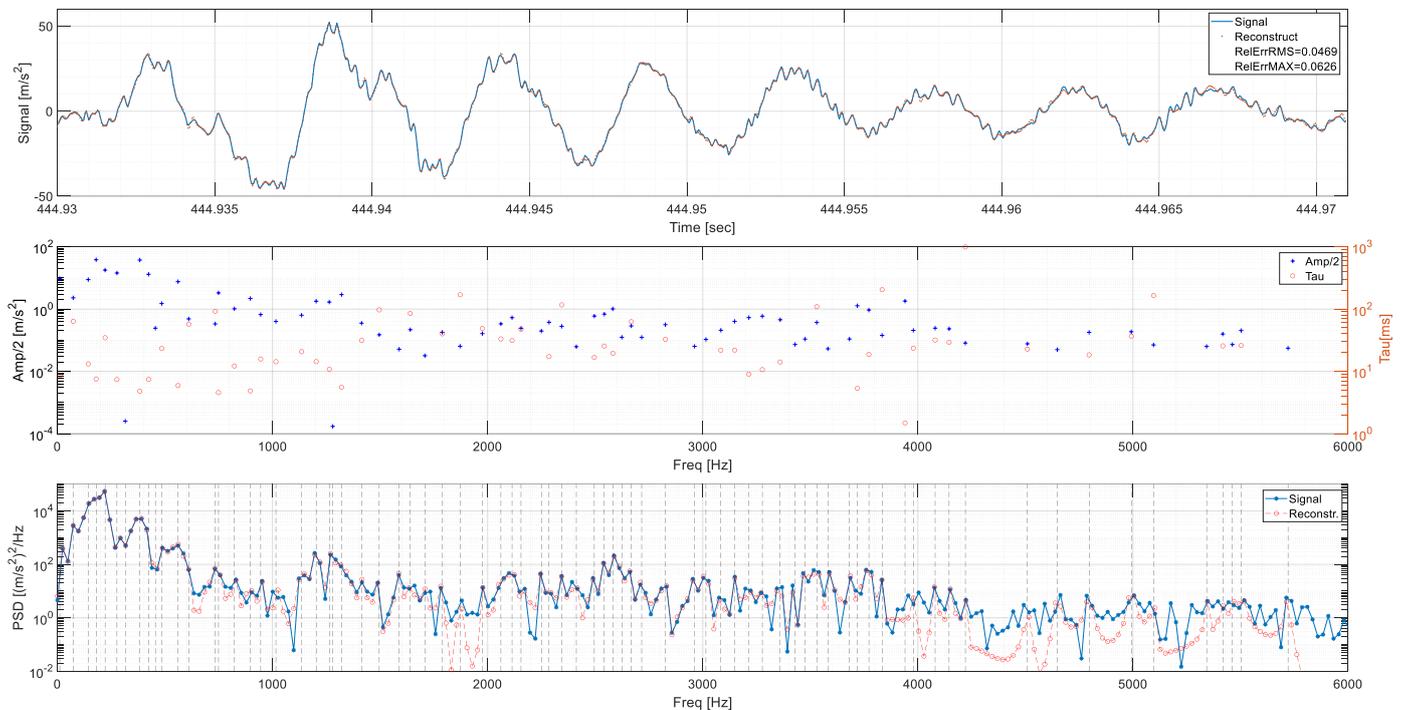

**Fig.13** HODMD modes extraction. Top: original and reconstructed signal, middle: amplitudes and time constants versus mode frequency, bottom: PSDs of the signal and the reconstructed signal, the dashed vertical lines representing the extracted HODMD frequencies.

## 3.3 Gliding HODMD on axel-end sensor

Based on previous description, we performed a gliding HODMD decomposition. The signal is sampled at $F_S$=25kHz, the whole analysed input signal length is L=$2^{13}$, the HODMD decomposition is done on N=$2^{10}$ samples (approximately 40ms) and the HODMD dimension used is d=500. The decomposition is performed for each sample and the following 1023 ones; each decomposition gives approximately 150-170 complex conjugate poles after optimal SVD truncation. **Fig.14** shows the input signal and the related extracted modes. We clearly observe a regularity in those mode frequencies appearing as horizontal dotted lines, as well as abrupt transients.

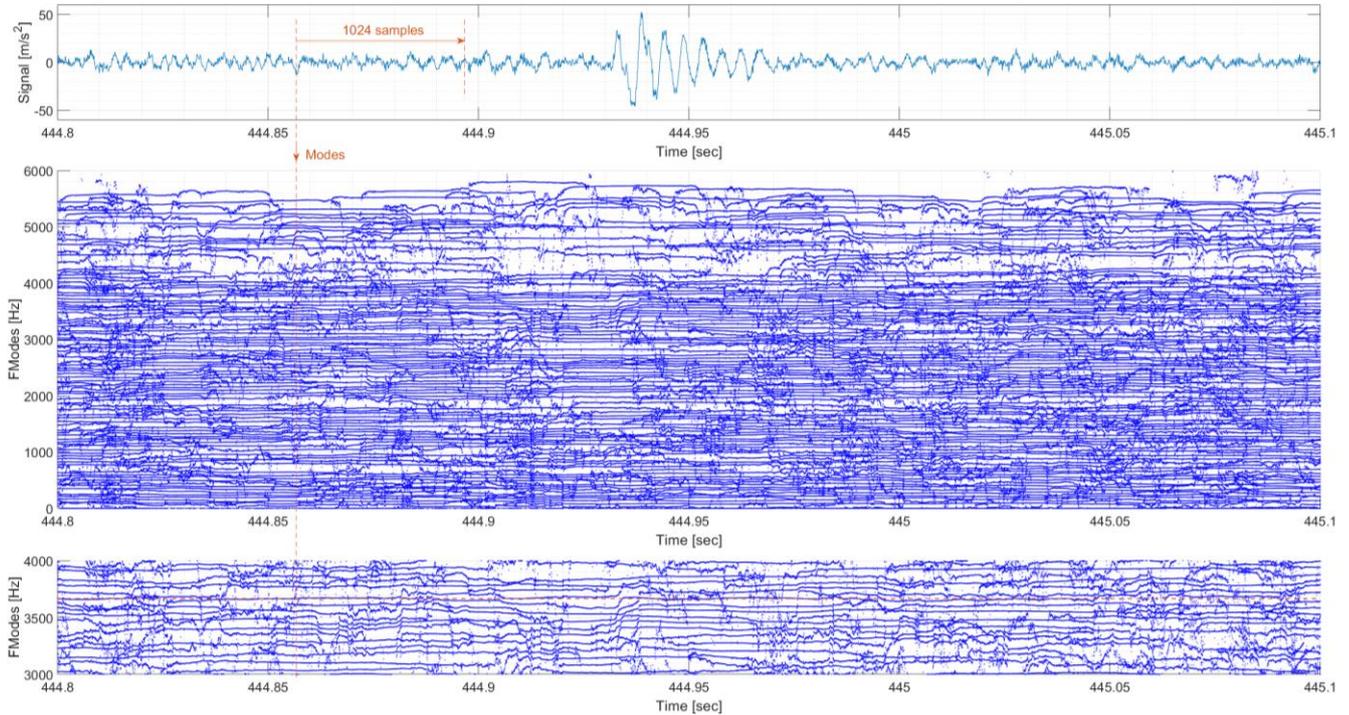

**Fig.14** HODMD modes extraction. Top: original signal, middle: whole frequency decomposition, bottom: zoom 3-4 kHz. The dashed vertical line over the 3 plots shows the mode extracted from the $2^{10}$ samples represented at the top plot. The horizontal line of 3670 Hz on the bottom figure shows that this frequency domain is heavily populated.

## 3.4 Kernel Density Spectrum during real operation

Both Gaussian and Lorentzian KDS are applied to the extracted modes of **Fig.14** and represented on figure **Fig.15** and **Fig.16** respectively. For each kernel the shape of the energy distribution of KDS and PSD are very similar. Most part of the energy is concentrated in the lower frequency range, roughly between DC and 600 Hz. We also observe humps and dips in the energy distribution at the same frequency regions with the difference that peaks emergent in a more distinct way. The key point is that its frequency resolution is highly improved and that, even if the spectrum looks like noisy, much sharper and distinct modes appear in the KDS.

To obtain a better representation, zooms in more restricted frequency range points out the very good spectral resolution. The HODMD/KDS resolutions are limited by the parameter h and the frequency resolution used compared to the PSD frequency bin. A comparison of two different parameter sets of h and the frequency resolution shows the ability to sharpen even further the spectrum depending on the required frequency resolution. This procedure has, of course, the drawback that if pushed too far it will become as noisy and difficult to interpret as a PSD. At the extreme limit it will finally reduce to a list of Diracs.

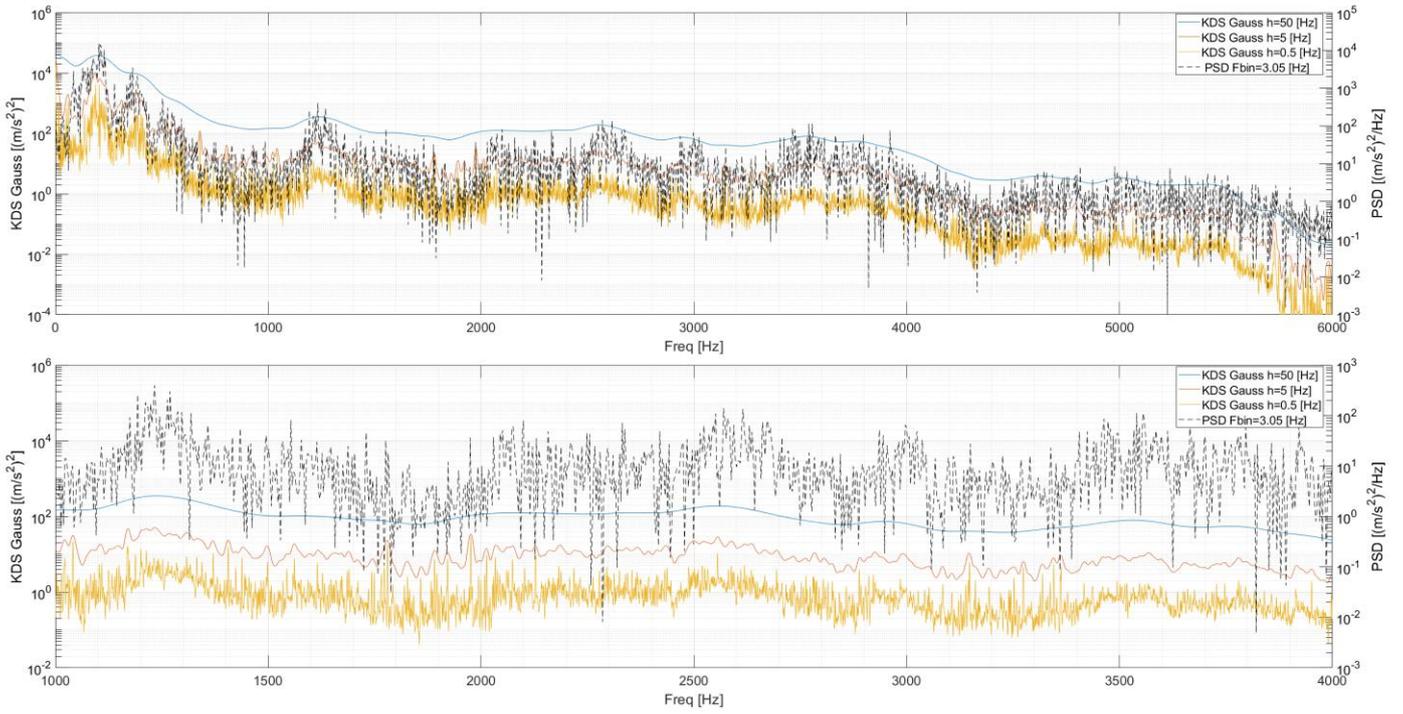

**Fig.15** Gaussian KDS with Amplitude² factor.

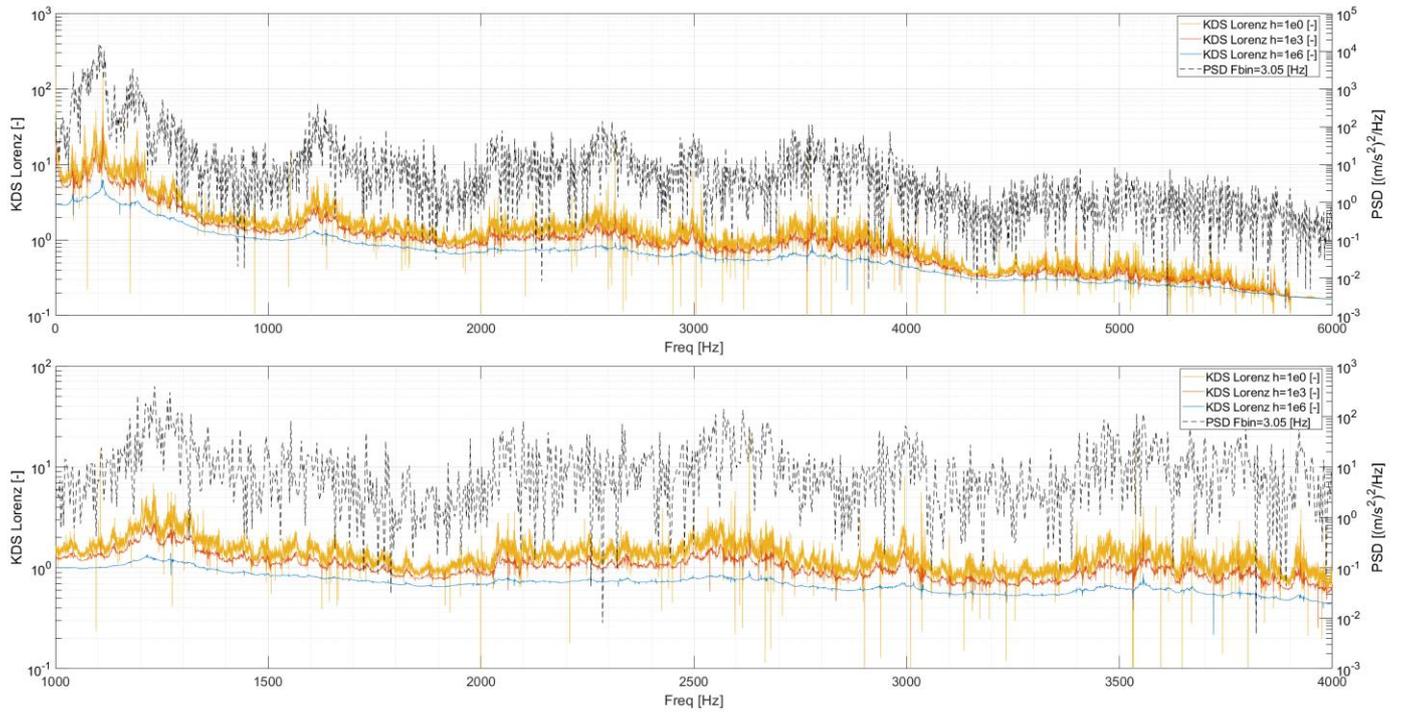

**Fig.16** Lorentzian KDS.

## **4. Conclusions**

In this paper we describe an alternative method on how to obtain a density or power-density spectrum which does not have some limitations as an FFT/PSD has: frequency resolution, spreading due to non-constant spectral amplitude (e.g. exponential decaying transients) and windowing leakage. We have shown that spectral analysis limits performed on real, transient and noisy vibration measurements can be overcome, opening the door for modal analysis of complex mechanical systems. This approach can easily be extended to non-stationary noisy signals of every kind.

**Glossary**

| | |
|---|---|
| DE | Differential Equations |
| ODE | Ordinary Differential Equation |
| PDE | Partially Differential Equations |
| FFT | Fast Fourier Transform |
| STFT | Short-Term Fourier Transform |
| PSD | Power Spectrum Density |
| DMD | Dynamic Mode Decomposition |
| HODMD | Higher Order Dynamic Mode Decomposition |
| KDE | Kernel Density Estimation |
| KDS | Kernel Density Spectrum |
| PCA | Principal Component Analysis |
| POD | Proper Orthogonal Decomposition |
| SVD | Singular Value Decomposition |

**Acknowledgments**

The author wishes to thank Vibro-Consult and Innosuisse for the possibility to work on this subject.

**Availability of data and materials**

The data is not available to the public due to confidentiality reasons.

**Authors' remark**

It may happen the author uses some nice formulated sentences taken from the references [1], [2] and [3]. The subject is so broad and deep, and details may be very complex, that it will be impossible to cover it within a paper and remember exactly where all used information were obtained from. So, if the author forgets citations or references, he is asking for indulgence.
Moreover, the author avoided too complex mathematical formulations and try to give a more intuitive feeling of this method. If the reader wants to understand the deepness of the subject there is no possible shortcuts to read and apply.

**Authors' contributions**

[1] developed this procedure and compiled this manuscript.
[2,3,6] contributed by delivering measurements used for this analysis.
[4,5,7] contributed by remarks, lectures and management activities.

**Competing interests**

The authors declare that they have no competing interests in this work.

**Author details**

[1,2,3,4,5]  Fachhochschule Nordwestschweiz, Institut für Sensorik und Elektronik,
Klosterzelgstrasse 2, CH-5210 Windisch

[6,7]  Vibro-Consult AG, Mess- und Schwingungstechnik
Stahlrain 6, CH-5200 Brugg

**Fundings**

This project was funded by Innosuisse project Nr. 37110.1 IP-ENG